\documentclass[conference,a4paper]{APSIPA2021}
\usepackage{multirow}
\usepackage{graphicx}
\usepackage{amsmath}
\usepackage[psamsfonts]{amssymb}
\usepackage{amsxtra}
\usepackage{threeparttable}
\usepackage{cite}
\usepackage{stfloats}
\usepackage[justification=centering]{caption}
\usepackage{subfigure}

\usepackage{geometry}
\usepackage{fancyhdr}

\newcommand{\reffig}[1]{Figure \ref{#1}}
\newcommand{\reftable}[1]{Table \ref{#1}}

\begin{document}


\title{A Policy-based Approach to the SpecAugment Method for Low Resource E2E ASR}

\author{%
\authorblockN{%
Rui Li\authorrefmark{1}\authorrefmark{4}, Guodong Ma\authorrefmark{1}\authorrefmark{4}, Dexin Zhao\authorrefmark{2}\authorrefmark{4}, Ranran Zeng\authorrefmark{2}, Xiaoyu Li\authorrefmark{2} and Hao Huang\authorrefmark{1}\authorrefmark{3}
}
\authorblockA{%
\authorrefmark{1} School of Information Science and Engineering, Xinjiang University, Urumqi, China 
}
\authorblockA{%
\authorrefmark{2} China Telecom Beijing Research Institute, Beijing, China
}
\authorblockA{%
\authorrefmark{3} Xinjiang Provincial Key Laboratory of Multi-lingual Information Technology, Urumqi, China \\
Corresponding author; E-mail: hwanghao@gmail.com
}
\authorblockA{%
\authorrefmark{4} Equal contributions
}
}

\maketitle

\begin{abstract}
SpecAugment is a very effective data augmentation method for both HMM and E2E-based automatic speech recognition (ASR) systems. Especially, it also works in low-resource scenarios. However, SpecAugment masks the spectrum of time or the frequency domain in a fixed augmentation policy, which may bring relatively less data diversity to the low-resource ASR. In this paper, we propose a policy-based SpecAugment (Policy-SpecAugment) method to alleviate the above problem. The idea is to use the augmentation-select policy and the augmentation-parameter changing policy to solve the fixed way. These policies are learned based on the loss of validation set, which is applied to the corresponding augmentation policies. It aims to encourage the model to learn more diverse data, which the model relatively requires. In experiments, we evaluate the effectiveness of our approach in low-resource scenarios, i.e., the 100 hours librispeech task. According to the results and analysis, we can see that the above issue can be obviously alleviated using our proposal. In addition, the experimental results show that, compared with the state-of-the-art SpecAugment, the proposed Policy-SpecAugment has a relative WER reduction of more than 10\% on the Test/Dev-clean set,  more than 5\% on the Test/Dev-other set, and an absolute WER reduction of more than 1\% on all test sets.
\end{abstract}

\section{Introduction}
Recently, end-to-end (E2E) automatic speech recognition (ASR) \cite{ctc-paper,RNNt-paper,LAS-paper,CTC_conformer,Speech-Transformer,Conformer} based neural networks have achieved a large improvements. 
Meanwhile, the E2E ASR simplifies the processing of system construction, which establishes a direct mapping from the acoustic feature sequences to the modeling unit sequences. 
With the emergence of E2E ASR, researchers \cite{SpecAugment,Phone_mask,pm_mmut,SpecAugment_large_dataset,mixspeech_paper,semantic_paper,jicheng_paper,yizhou_paper,decouple_pronunciation,hybrid_modeling_units,augmentation_paper1,augmentation_paper2,auditory-based-paper,Phone-informed-paper,importanaug-paper,task-aug-kinder-paper}  explore different E2E ASR scenarios and partly focus on the data augmentation and training strategy due to the nature of data-hungry and easy over-fitting. However, most of the existing work on data augmentation technology is just to explore a fixed way to bring more abundant data to the model. For example, the state-of-the-art SpecAugment method \cite{SpecAugment} uses three spectrum disturbance strategies in a fixed augmentation policy for input speech spectrum features. 
Under a certain amount of training data, the fixed augmentation policy may tend to be stable faster so that it will not bring too much data diversity in a model learning stage. Therefore, we believe that, with the use of SpecAugment, the information brought to the model can be enriched. In addition, when the model is in different learning states, it may more need to learn the data applied to different the combinations of augmentation policy. But the masking strategies of SpecAugment are completely random which is not related to model state.

Based on the above discussions, we propose a policy-based SpecAugment to improve the performance of low resource end-to-end (E2E) ASR systems, which is named Policy-SpecAugment.
In our proposed Policy-SpecAugment method, we will calculate the loss value of the validation set under the action of each augmentation strategies in the model learning stage.  
The loss value can be represented the fitting degree of the model to the corresponding augmentation strategies at the previous epoch trained model, so as to reflect which augmentation strategy should be studied at the current model training epoch. Then, the losses will be used to calculate the probabilities of the augmentation-select policy and the factor of the augmentation-parameter changing policy, which aims to encourage the augmentation method to produce the various data needed by ASR model. The specific details will be introduced in section \ref{sec:policy-specaugment}. For fair comparison, the augmentation strategies we used are consistent with SpecAugment, including time masking, frequency masking and time warping. We will briefly show the three classic and effective augmentation strategies in Section \ref{sec:augmentation-strategies}. In the 100 hours librispeech task \cite{Librispeech-paper}, we use ESPnet1 \cite{espnet} to confirm our Policy-SpecAugment. The experimental results show that, compared with the state-of-the-art SpecAugment, our proposed Policy-SpecAugment has a relative increase of more than 10\% on the test/dev-clean set, a relative increase of more than 5\% on the test/dev-other set, and an absolute increase of more than 1\% on all test sets. 

The paper is organized as follows. Section \ref{sec:relate-works} is to review the prior related works. In Section \ref{sec:augmentation-strategies}, we will briefly show the three data augmentation methods in SpecAugment. And Section \ref{sec:policy-specaugment} presents the proposed policy-based SpecAugment method in detail. In addition, Section \ref{sec:exp_res} describes the experiment setups and results. After that, we perform further analysis in Section \ref{sec:analysis} and conclude in Section \ref{sec:conclusion}.

\section{RELATED WORKS}
\label{sec:relate-works}
Data augmentation is a method of increasing the diversity of training data\cite{cvsurvey,nlpsurvey,srsurvey} to prevent the model over-fitting. Currently, there are many data augmentation methods to improve ASR system performance. In \cite{speed-perturb}, audio speed perturbation was proposed, which aims to augment the varied speed data for the model. In \cite{virtualrooms}, room impulse responses were proposed to simulate far-field data.
In \cite{synthetic}, speech synthesis methods are used to augment the data. In addition, Our recent proposed Phone Masking Training (PMT) \cite{Phone_mask} alleviates the impact of the phonetic reduction in Uyghur ASR by simulating phonetic reduction data. Then, the state-of-the-Art SpecAugment \cite{SpecAugment} has proven to be a very effective method, which operates on the input speech spectrum features of the ASR model using the three augmentation strategies, including time masking, frequency masking and time warping. 

However, the data augmentation methods mentioned above both enrich the training data in a fixed augmentation policy, which does not choose the diversity of data based on the state of the model learning. We believe that the fixed way will reduce the data diversity brought by the data augmentation method. To alleviate this issue, the recent SapAugment  \cite{SapAugment} proposes to use the loss value of the training sample to make a selection for the augmentation operation at the corresponding training sample. Their intuitions are that perturbing a hard sample with a strong augmentation may also make it too hard to learn from, and a sample with low training loss should be perturbed by a stronger augmentation to provide more robustness to a variety of conditions.  \cite{SapAugment} shows an obvious performance advantage than SpecAugment. But, more strategies used in SapAugment (such as CutMix \cite{cutmix} etc.) are introduced compare with SpecAugment. In addition, we believe that the strategy selection should be more focused on the strategy rather than data samples. 


\begin{figure*}[htp]
\begin{center}
\includegraphics[width=6in]{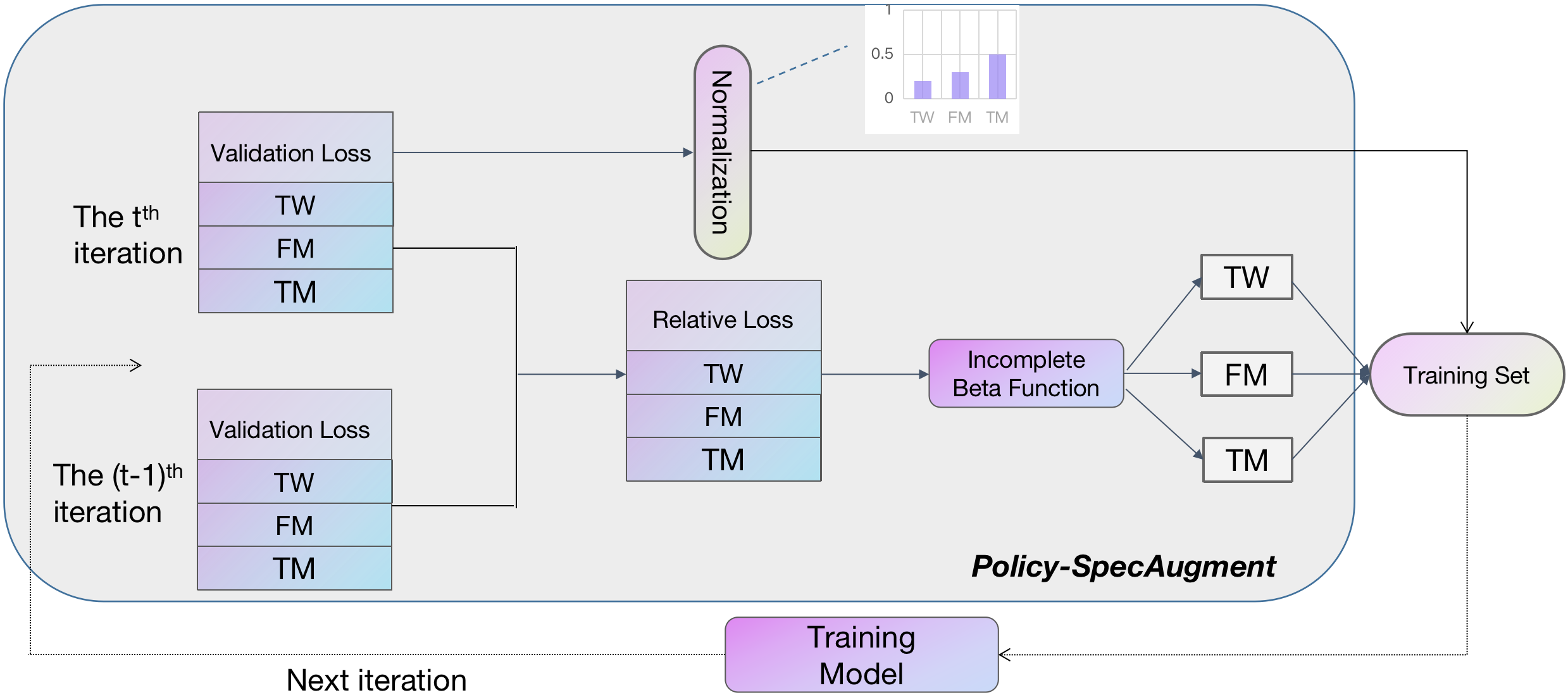}
\caption{Policy-SpecAugment augmentation method, where TW denotes Time Warping, FM presents Frequency Masking and TM refers to Time Masking.}
\label{model-structure}
\end{center}
\end{figure*}



\section{AUGMENTATION STRATEGY}
\label{sec:augmentation-strategies}
Our motivation is to construct a policy that allows the model to fully learn the benefits of different data augmentation methods as it is trained. For fair comparison, we use the same data augmentation methods as SpecAugment to our proposed Policy-SpecAugment. The following is a brief introduction to three data augmentation strategies used in SpecAugment. Please refer to \cite{SpecAugment} for the details.

\subsection{Time Masking}
Time masking is to mask the input features in the time domain. The masked size is first chosen from a uniform distribution from 0 to the time mask parameter \emph{T}.

\subsection{Frequency Masking}
Frequency masking masks the input spectrum features in the frequency domain. The masked part is first chosen from a uniform distribution from 0 to the frequency mask parameter \emph{F}.

\subsection{Time Warping}
About time warping, given an input speech spectrum features, a random point along the horizontal line passing through the center of the spectrum is to be warped either to the left or right by a certain distance chosen from a uniform distribution. 

\section{POLICY-BASED SPECAUGMENT}
\label{sec:policy-specaugment}
\subsection{Random-SpecAugment}
Before the introduction of the Policy-SpecAugment, we will show the Random-SpecAugment. Briefly, each training sample were applied to only one augmentation strategy in equal random way. Though Random-SpecAugment with the very simple idea is worsened than SpecAugment in experiments, it outperform obviously the baseline without any augmentation method and motivates us to propose the Policy-SpecAugment.

\subsection{Prob-SpecAugment}
Based on the Random-SpecAugment, to fix the equal random way, we propose a probability-based SpecAugment (Prob-SpecAugment). The idea is that, at each iteration of the training phase, we will calculate \emph{N} loss values ($Loss_{i}$, ${i = 1 .. N}$) of valid set applied to the corresponding augmentation technology, where $N$ displays the total number of augmentation strategies.
We consider the normalized loss as the probability distribution of the augmentation strategies, as shown in Equation \eqref{score}. Then these probabilities are used to replace the equal probability of Random-SpecAugment. It represents that each sample is augmented by only one augmentation policy. 

\begin{equation}
    \label{score}
    \mathcal{P}_{i} = \frac{\mathcal{L}oss_{i}}{\sum_{i=1}^N\mathcal{L}oss_{i}}
\end{equation}


\begin{table*}[t]
  \begin{center}

  \caption{WERs (\%) of on LibriSpeech 100 hours task. The $0^{th}$ system is the baseline model without any data augmentation; SpecAugment is only used to the ${1^{th}}$ system; Random-SpecAugment is applied to the ${2^{th}}$ system; The ${3^{th}}$ system  use the Prob-SpecAugment; The ${4^{th}}$ system apply Prob-SpecAugment plus the IBF Param change policy of Policy-SpecAugmetn; The ${5^{th}}$ system is based on the ${4^{th}}$ system to add SpecAugment; Finally, the ${6^{th}}$ system is only under the function of Policy-SpecAugment. In addition, the numbers in brackets indicate the absolute WER reduction comparing with SpecAugment system.}
  \renewcommand\tabcolsep{10.0pt}
 
  \begin{tabular}{l|l|c|c|c|c}
  \hline
        Index & SYSTEM & Test-clean & Test-other & Dev-clean & Dev-other  \\ \hline      
        0 & Baseline & 12.1 (-1.6) & 32.7 (-9.5) & 11.3 (-1.5) & ~32 (-9.7)  \\ \hline
        1 & SpecAugment & 10.5 (~0.0) & 23.2 (~0.0) & ~9.8 (~0.0) & 22.3 (~0.0) \\ \hline
        2 & Random-SpecAugment & 10.6 (-0.1) & 26.5 (-3.3) & 10.3 (-0.5) & 25.3 (-3~)  \\ \hline
        3 & Prob-SpecAugment & 10.1 (+0.4) & 24.4 (-1.2) & ~9.1 (+0.7) & 24.4 (-2.1)  \\ \hline
        4 & Prob-SpecAugment + IBF Param change & ~9.7 (+0.8) & 23.0 (+0.2) & ~8.9 (+0.9) & 22.2 (+0.1)  \\ \hline
        5 & Prob-SpecAugment + IBF Param change + SpecAugment & 10.1 (+0.4) & 22.2 (+1.0) & ~8.9 (+0.9) & 21.4 (+0.9)  \\ \hline
        6 & Policy-SpecAugment & ~\textbf{9.1 (+1.4)} & \textbf{21.5 (+1.7)} & ~\textbf{8.3 (+1.5)} & ~\textbf{21 (+1.3)} \\ \hline
  \end{tabular}
  \label{result}
   \end{center}
\end{table*}

\subsection{Policy-SpecAugment}
\label{sec:sec:policy-specaugmet}

Based on but unlike Prob-SpecAugment, these probabilities from Equation \eqref{score} are used to the function on-off of these augmentation strategies for each sample of the training set. It means that each sample is augmented by at least one strategy and at most three. The overview of our proposed Policy-SpecAugment is shown in Figure \ref{model-structure}. 

In addition, we can seen from Figure \ref{model-structure}, to increase more diversity of the augmentation strategies, we follow \cite{SapAugment} to change the parameters of the augmentation strategies by using the incomplete beta function \cite{ibf_paper}. Unlike \cite{SapAugment}, we use the relative loss values, which calculated from the current iteration model and the previous iteration, to get the change factor, as shown in Equation \eqref{relative_loss} and Equation \eqref{ibf}. We 
believe that the relative loss can be well to represent the model learning degree to the corresponding augmentation strategies. If the relative loss value is low, it is allowed to display that the model is not sensitive to that augmentation strategy in the previous model state. Then, this strategy should be changed and encouraged to the current model training epoch.  

\begin{equation}
    \label{relative_loss}
    Relative_{i,j} = \begin{cases}
    \frac{Loss_{i,j-1} - Loss_{i,j}}{Loss_{i,j-1}}, & Loss_{i,j} < Loss_{i,j-1} \\
    \frac{Loss_{i,j} - Loss_{i,j-1}}{Loss_{i,j}},&Loss_{i,j} \ge Loss_{i,j-1}
    \end{cases}
\end{equation}
\begin{equation}
    \label{ibf}
    \lambda_{i,j} = 1 - {IBF}(a,b,Relative_{i,j})
\end{equation}
Where \emph{j} represents the $\emph{j}^{th}$ iteration, $\emph{Relative}_{i,j}$ means the relative loss at the $j^{th}$ iteration of the $i^{th}$ augmentation strategy. a, b are two hyper-parameters, we set $a=0.6,b=4.4$. $\lambda_{i,j}$ is the factor used to change the parameters of the augmentation strategy. Especially, in the start of model leaning, we will use the Random-SpecAugment to get the first epoch model. And we will use the first model and  zero loss to get the relative loss of valid set which are applied to the corresponding augmentation strategy. The parameter changing formula follow \cite{SapAugment}, as shown in \reftable{parameter_change}.

\begin{table}[h]
  \caption{The parameter changing mapping.}
  \renewcommand\tabcolsep{5.0pt}
  \centerline{}
  \label{parameter_change}
  \centering
  \begin{tabular}{l|c|c}
    \hline
    {{Augmentation Parameter}}  & {{Range}} & {{Mapping from $\lambda$}} \\
    \hline
    $\rho_0$ (Time Warping) & [0.2, 0.6] & $\rho_0$ = 0.2 + 0.4$\lambda$  \\
    $N_{masks}$ (Time Masking)  & \{2, 3, 4, 5, 6\} & $N_{masks}$ =  $\lfloor2 + 4\lambda\rfloor$ \\
    $N_{masks}$ (Time Masking)  & \{2, 3, 4, 5, 6\} & $N_{masks}$ =  $\lfloor2 + 4\lambda\rfloor$ \\
    \hline
  \end{tabular}
\end{table}

Our intuition of the proposed Policy-SpecAugment is to encourage the model to use the model learning state to learn the data applied to the more suitable strategy, so that the model can better learn the performance improvement that each augmentation strategy encourages.


\begin{table}[h]
  \caption{The datasets }
  \renewcommand\tabcolsep{5.0pt}
  \centerline{}
  \label{data-description}
  \centering
  \begin{tabular}{l|l|cc}
    \hline
    &{{Data Type}}  & {{Duration (Hours)}} & {{Domain}} \\
    \hline
    &Train-100 & 100 & Read/Clear \\
    English &Test/Dev-clean & 5.4 & Read/Clear \\
    &Test-other  & 5.1 & Read/Noisy \\
    &Dev-other  & 5.3 & Read/Noisy\\
    \hline
  \end{tabular}
\end{table}

\section{EXPERIMENTS AND RESULTS}
\label{sec:exp_res}

\subsection{Data Description}
The proposed method is evaluated on Librispeech \cite{Librispeech-paper} 100 hours speech recognition low resource task, and test on the standard test-clean/other and dev-clean/other. The details of the datasets are shown in Table~\ref{data-description}, which is available at OpenSLR website\footnote{https://www.openslr.org/12/}.

\subsection{Experiment Setup}
Following \cite{recent-conformer}, we use  80-dimensional logmel spectral energies plus 3 extra features for pitch information as acoustic features input. Following \cite{Phone_mask,recent-conformer}, the trade-off CTC weight in model training is set to 0.3 and in decoding is set 0.5. For the other configurations, we use a similar setup in the ESPnet official configuration for the Librispeech 100 task\footnote{https://github.com/espnet/espnet/tree/master/egs/librispeech100/asr1}.
All the E2E models (Encoders = 12, Decoders = 6, Aheads = 4, $d^{att}$= 256) are trained by using  ESPnet1 \cite{espnet} on only one 1080 Ti GPU with 16GB
memory over all experiments. No external language models are used. In addition, The output tokens are 300 BPE tokens produced by unigram model using sentencepiece \cite{SentencePiece} in all tasks.

\begin{figure*}[htp]
\begin{center}
\subfigure[Baseline without data augmentation]{\includegraphics[scale=0.15]{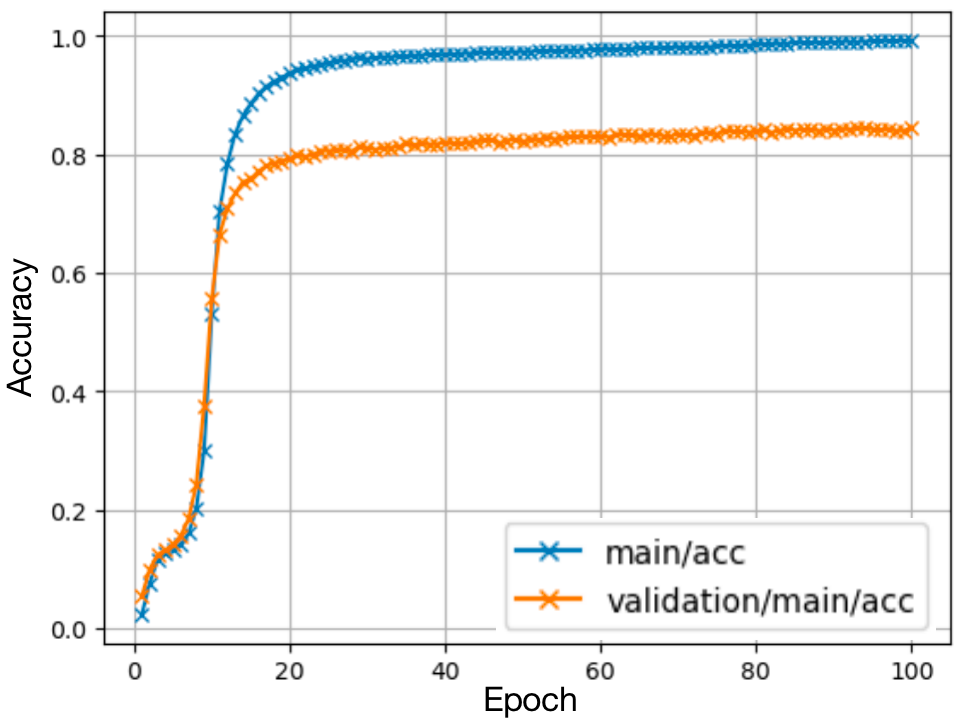}}
\hspace{5mm}
\subfigure[SpecAugment]{\includegraphics[scale=0.15]{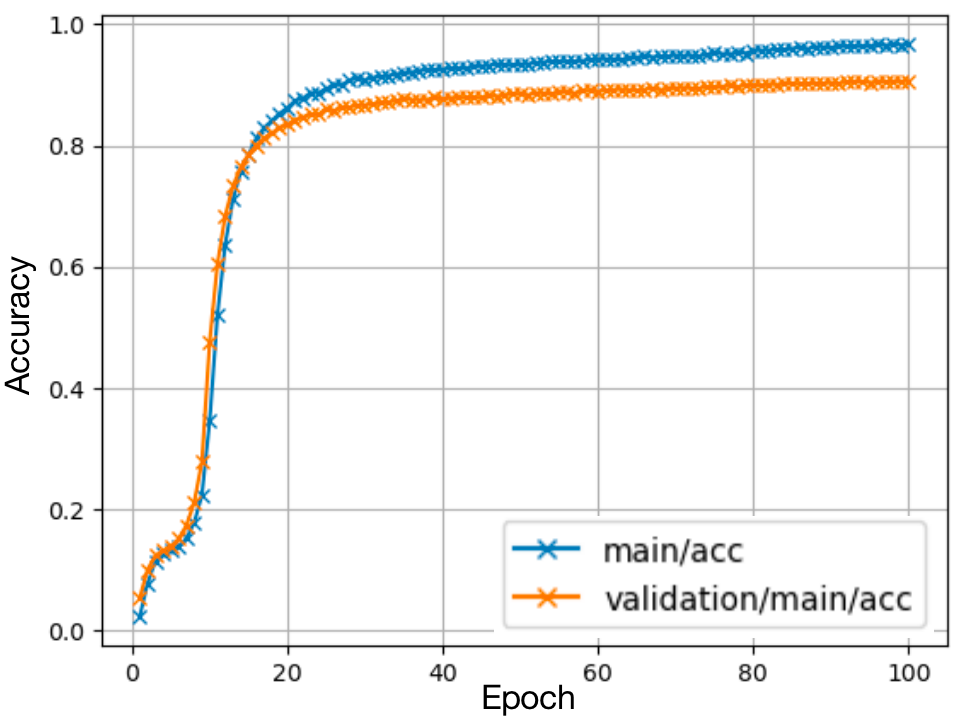}}
\hspace{5mm}
\subfigure[Policy-SpecAugment]{\includegraphics[scale=0.15]{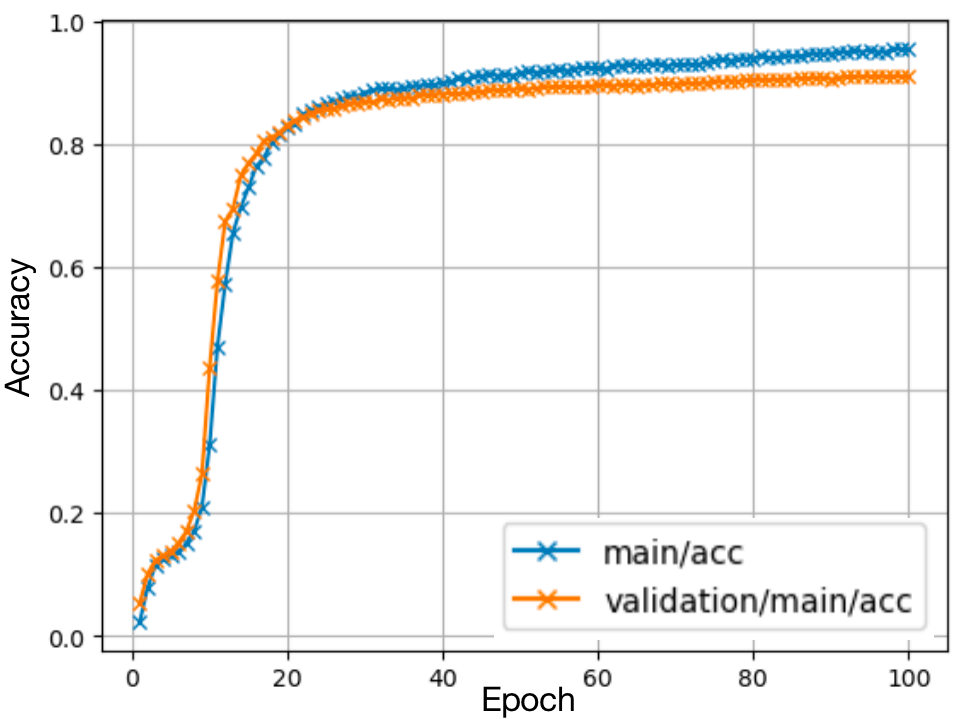}}
\caption{Accuracy on LibriSpeech 100 hours}
\label{acc}
\end{center}
\end{figure*}

\subsection{Results}

\reftable{result} reports the results. 
The $0^{th}$ system is the baseline model without any data augmentation. The $1^{th}$ system apply only SpecAugment. Comparing the experimental results of these two systems, we can observe that the $1^{th}$ system show a significant improvement both on the test/dev-clean set and the test/dev-other set. This means that SpecAugment greatly improves the performance of ASR task in low-resource scenarios. Further, the $2^{th}$ system uses Random-SpecAugment which apply only one random augmentaton strategy for a training sample. Comparing the experimental results between the system $1^{th}$ and $2^{th}$, we can see that the former is better. But, the system $2^{th}$ outperforms obviously the baseline system. The reason might be that the SpecAugment uses three augmentation strategies simultaneously for a training sample.   
We think this results is interesting, is it really because of the number of augmentation strategies used, or is it because the model has not learned enough about the benefits of a particular augmentation strategies? 

Inspired by the above results, we experiment Prob-SpecAugment, which replaces the equal probability of Random-SpecAugment with Equation \eqref{score}, in the system $3^{th}$.  
Comparing the results of Random-SpecAugment, Prob-SpecAugment system has an obvious WER improvement. Especially, Prob-SpecAugment is better than SpecAugment in Test/Dev-clean and further reduce the WER gap in Test/Dev-other. Based on Prob-SpecAugment, we apply incomplete beta function to change the parameters of the augmentation strategies according to the relative losses (please refer to Section \ref{sec:policy-specaugment} for details). As shown in the $4^{th}$ system of \reftable{result}, it outperforms the SpecAugment in all test sets. 
Although the $4^{th}$ system show an well WER on all test sets comparing with the SpecAugment, it has a little improvement in Test/Dev-other sets. We guess that it might be related to the number of augmentation strategies in each training sample, since the Prob-SpecAugment only choose one of the three augmentation methods to be used on a training sample based on the probability. To verify this idea, we combine system 4 and SpecAugment in the $5^{th}$ system.
We can see from the \reftable{result} that System $4^{th}$ show an obvious WER improvement in Test/Dev-other sets. 
This shows that the number of augmentation strategies is related to the robustness of the system. 

But, the SpecAugment masks the spectrum of time or the frequency domain in a fixed augmentation, which may be increase the learning pressure of the model in low resource ASR task. To alleviate it, we propose the Policy-SpecAugment (please refer to Section \ref{sec:sec:policy-specaugmet} for the details).  
We can see from \reftable{result} that, comparing with SpecAugment, the proposed Policy-Specaugment has a relative WER reduction of more than 10\% on the test/dev-clean set, more than 5\% on the test/dev-other set, and an absolute WER reduction of more than 1\% on all test sets. Our intuition  of the proposed Policy-SpecAugment mentioned by the above is to encourage the model to use the model learning state to learn the data applied to the suitable strategy, so that the model can better learn the performance improvement that each augmentation strategies encourages.

\section{ANALYSIS}
\label{sec:analysis}
\subsection{Accuracy}
\reffig{acc} shows the accuracy of the baseline, SpecAugment and our proposed Policy-SpecAugment. We can observe that the baseline system leads to the lowest accuracy of the validation set and a largest loss gap between the accuracy of the training and validation sets. This loss gap may show the degree of over-fitting in some certain.  When SpecAugment is used, the accuracy of the validation set is improved and the loss gap is reduced. it indicates that the robustness of the model is improved by using SpecAugment. Further, when our proposed Policy-SpecAugment is used, the loss gap is further reduced, which proves that our Policy-SpecAugment can prevent over-fitting better than SpecAugment. And then, our method makes the model more robust and gets the better results.

In general, our proposed Policy-SpecAugment will bring more data diversity for model learning than SpecAugment in low resource scenarios, which is consistent with our motivation.

\subsection{Probability}
\begin{figure}[htp]
\begin{center}
\includegraphics[scale=0.35]{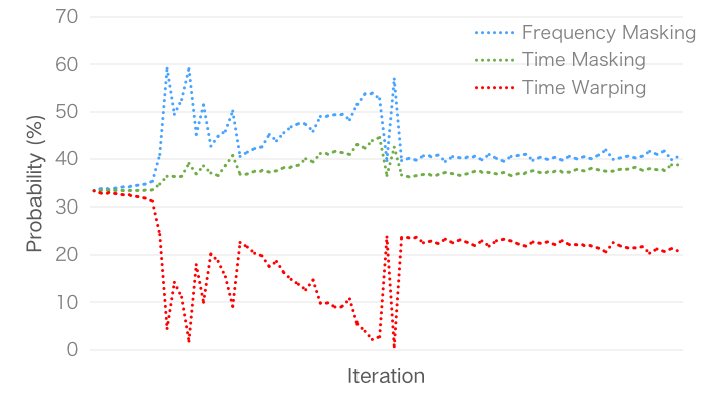}
\caption{Probability of Augmentation}
\label{prob}
\end{center}
\end{figure}

Meanwhile, we also analyse the probabilities of each data augmentation strategies, which include Frequency Masking (FM), Time Masking (TM) and Time Warping (TW), in our model training stage, as shown in \reffig{prob}. Importantly, the probabilities are initialized in the equal way. The probability of frequency masking is always the largest, which might bring the more data diversity than other two policies in the modeling training. Other, the probability of time warping show an low value. In practical application of SpecAugment, the time warping plays less role than the other two. For example, in WeNet toolkit \cite{wenet_paper}, the default SpecAugment configuration \footnote{https://github.com/wenet-e2e/wenet/blob/main/examples/librispeech/s0/\\conf/train\_conformer.yaml} has no time warping strategies. 

In other words, this results have proved our intuition, i.e. the model can better learn the performance improvement that each augmentation strategies wants to lead.

\subsection{Relative Loss and 1 - IBF value}
\begin{figure}[htp]
\begin{center}
\subfigure[Relative Loss]{\includegraphics[scale=0.30]{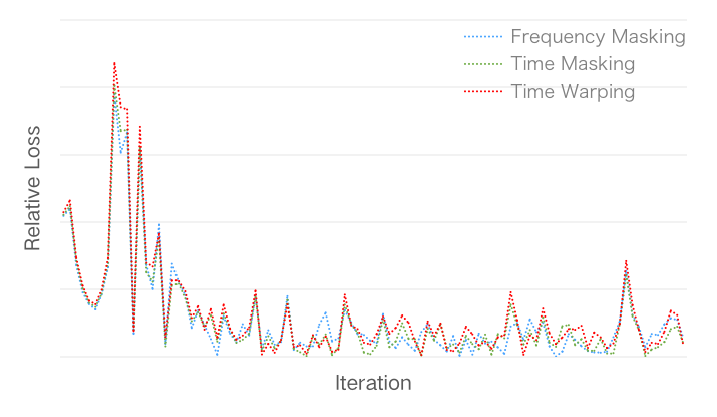}} \\
\subfigure[1-IBF value]{\includegraphics[scale=0.30]{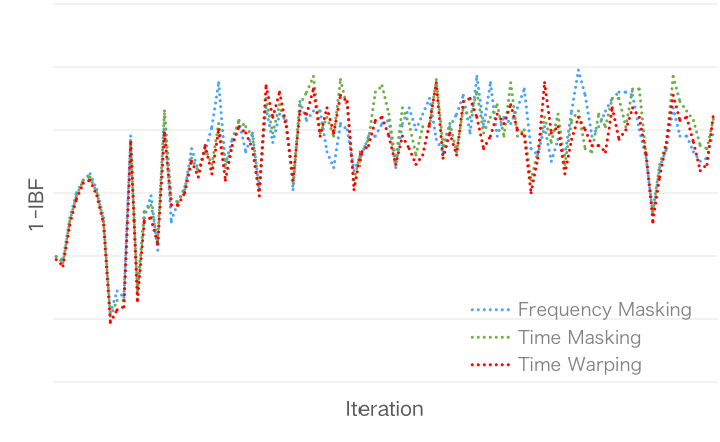}}
\caption{Parameters of Augmentation }
\label{relativeloss}
\end{center}
\end{figure}

Finally, we conduct the an analysis about the change trend of the relative loss and 1 - IBF value, which include Frequency Masking (FM), Time Masking (TM) and Time Warping (TW). The results is shown in \reftable{relativeloss}. From \reffig{relativeloss}, we will change the parameter to produce the more data to next model learning stage by 1 - IBF value, when the relative loss changes. If the relative loss of a sample is small, it means that the model may not be able to learn more information from this sample. Therefore, we will increase the amplification intensity of this sample to encourage the model to explore deeper information of this sample. This is also to consider the learning ability of the model at different states.

In this way, we hope to encourage the model to learn robustness in different data produced by our proposed.

\section{CONCLUSIONS}
\label{sec:conclusion}
In this paper, we propose Policy-SpecAugment to use the augmentation-select policy and the augmentation-parameter changing policy to solve the fixed way of the state-of-the-art SpecAugment. It may prevent over-fitting and bring more data diversity in low-resource ASR scenarios. The augmentation-select policy is to adjust the data augmentation strategies according to the probabilities. These probabilities are calculated with validation set applied to the corresponding augmentation strategies during model training stage. 
The augmentation-parameter changing policy aims to dynamically change the parameters of each augmentation strategies according to the relative loss. Through the above, it is allowed to encourage the model to use the model learning state to learn the data applied to the suitable strategy, so that the model can better learn the performance improvement that each augmentation strategies wants to bring. 
We have carried out low resource ASR task on librispeech 100 hours. And the results show that the proposed method obviously improves the recognition performance. Meanwhile, according to the analysis, our proposed Policy-SpecAugment is helpful in improving the low resource English E2E ASR.

\vfill\pagebreak
\section{ACKNOWLEDGEMENTS}
This work was supported by the National Key R\&D Program of China (2020AAA0107902), Opening Project of Key Laboratory of Xinjiang, China (2020D04047), and Natural Science Foundation of China (61663044, 61761041).



\bibliographystyle{IEEEbib}
\bibliography{refs}

\end{document}